# CellSense: A Probabilistic RSSI-based GSM Positioning System

Mohamed Ibrahim
Wireless Intelligent Networks Center (WINC)
Nile University
Smart Village, Egypt
Email: m.ibrahim@nileu.edu.eg

Moustafa Youssef
Wireless Intelligent Networks Center (WINC)
Nile University
Smart Village, Egypt
Email: mayoussef@nileu.edu.eg

***Abstract*—Context-aware applications have been gaining huge interest in the last few years. With cell phones becoming ubiquitous computing devices, cell phone localization has become an important research problem. In this paper, we present CellSense, a probabilistic RSSI-based fingerprinting location determination system for GSM phones. We discuss the challenges of implementing a probabilistic fingerprinting localization technique in GSM networks and present the details of the CellSense system and how it addresses the challenges. To evaluate our proposed system, we implemented CellSense on Android-based phones. Results for two different testbeds, representing urban and rural environments, show that CellSense provides at least 23.8% enhancement in accuracy in rural areas and at least 86.4% in urban areas compared to other RSSI-based GSM localization systems. This comes with a minimal increase in computational requirements. We also evaluate the effect of changing the different system parameters on the accuracy-complexity tradeoff.**

## I. Introduction

As cell phones become more ubiquitous in our daily lives, the need for context-aware applications increases. One of the main context information is location that enables a wide set of cell phone applications including navigation, location-aware social networking, and security applications. GPS is considered one of the most well known localization techniques [1]. However, GPS is not available in many cell phones, requires direct line of sight to the satellites, and consumes a lot of energy. Therefore, research for other techniques for obtaining cell phones' location has gained momentum fueled by both the user needs for location-aware applications and government requirements, e.g. FCC [2]. City-wide WiFi-based localization for cellular phones has been investigated in [3], [4] and commercial products are currently available [5]. However, WiFi chips, similar to GPS, are not available in many cell phones and not all cities in the world contain sufficient WiFi coverage to obtain ubiquitous localization. Similarly, using augmented sensors in the cell phones, e.g. accelerometers and compasses, for localization have been proposed in [6]–[8]. However, these sensors are still not widely used in many phones. On the other hand, GSM-based localization, by definition, is available on all GSM-based cell phones, which presents 80-85% of today's cell phones [9], works all over the world, and consumes minimal energy in addition to the standard cell phone operation. Many research work have addressed the problem of GSM localization [2], [4], [10], [11], including time-based systems, angle-of-arrival based systems, and received signal strength indicator (RSSI) based systems. Only recently, with the advances in cell phones, GSM-based localization systems have been implemented [4], [10], [11]. These systems are mainly RSSI-based as RSSI information is easily available to the user applications. Since RSSI is a complex function of distance, due to the noisy wireless channel, RSSI-based systems usually require building an RF fingerprint of the area of interest [4], [10], [11]. A fingerprint stores information about the RSSI received from different base stations at different locations in the area of interest. This is usually constructed once in an offline phase. During the tracking phase, the received RSSI at an unknown location is compared to the RSSI signatures in the fingerprint and the closest location in the fingerprint is returned as the estimated location. Constructing the fingerprint is a time consuming process. However, this is typically done in a process called war driving, where cars scan the street of a city to map it. Current commercial systems, such as Skyhook, Google's MyLocation and StreeView services already perform scanning for other purposes. Therefore, constructing the fingerprint for GSM localization can be *piggybacked on these systems* ***without extra overhead***.

In this paper, we propose *CellSense*, a **probabilistic** fingerprinting based techniques for GSM localization. Unlike the current fingerprinting techniques for GSM phones that uses deterministic techniques for estimating the location of cell phones [10], [11], *CellSense* probabilistic technique provides more accurate localization. However, constructing a probabilistic fingerprint is challenging, as we need to stand at each fingerprint location for a certain amount of time to construct the signal strength histogram. This adds significantly to the overhead of the fingerprint construction process. *CellSense* addresses this challenge by using gridding, where the area of interest in divided into a grid and the histogram is constructed for each grid cell. This, not only removes the extra overhead of standing at each location for a certain time, but also helps in increasing the scalability of the technique as the fingerprint size can be reduced arbitrarily by increasing the grid cell size. To evaluate *CellSense*, we implemented it on Android-enabled cell phones and compare its performance to other deterministic fingerprinting techniques, model based techniques, and Google's MyLocation service under two different testbeds representing rural and urban environments. We also study the effect of the different parameters on the performance of *CellSense*.

The rest of the paper is organized as follows: Section II discusses the different techniques for RSSI-based localization in GSM networks. In Section III, we present our *CellSense* system. Section IV presents the performance evaluation of our system. Finally, Section V concludes the paper and gives directions for future work.

## II. Background

This section presents a brief background on the current RSSI-based techniques for GSM localization that we use for comparison with *CellSense* including: cell-ID based techniques, deterministic fingerprinting techniques, and modeling-based techniques.

### A. Cell-ID based Techniques

Cell-ID based techniques, e.g. Google's MyLocation [12], do not use RSSI explicitly, but rather estimate the cell phone location as the location of the cell tower the phone is currently associated with. This is usually the cell tower with the strongest RSSI. Such techniques require a database of cell towers' locations and provide an efficient, though coarse grained localization method.

### B. Deterministic Fingerprinting Techniques

Fingerprinting based techniques store the RSSI signature of cell towers at different locations in the area of interest in a database during an offline phase. This database is searched during the tracking phase for the closest location in the RSSI space to the unknown location. Fingerprints are usually constructed by war driving, where a car drives the area of interest continuously scanning for cell towers and recording the cell tower ID, RSSI, and GPS location.

Current fingerprinting techniques for GSM localization use only determinist techniques [10], [11]. For example, each location in the fingerprint of [10] stores a vector representing the RSSI value from each cell tower heard at this location. During the tracking phase, the K-Nearest Neighbors (KNN) classification algorithm is used, where the RSSI vector at an unknown location is compared to the vectors stored in the fingerprint and the K-closest fingerprint locations, in terms of Euclidian distance in RSSI space, to the unknown vector are averaged as the estimated location. Deterministic fingerprinting techniques require searching a larger database than cell-ID based techniques but provide higher accuracy. Note that the overhead of constructing the fingerprint is the same as constructing the cell ID database as both require war driving.

### C. Modeling-based Techniques

Modeling-based techniques try to capture the relation between signal strength and distance using a model. For example, the work in [10] uses a Gaussian process to capture this relation assuming that the received signal strength $y_i$ at location $x_i$ is $y_i = f(x_i) + \epsilon_i$ Where $\epsilon_i$ is zero mean, additive Gaussian noise with known variance $\sigma_n^2$.

A Gaussian process (GP) estimates posterior distributions over functions $f$ from a training data $D$ (fingerprint). These distributions are represented non-parametrically, in terms of the training points. A key idea underlying GP's is the requirement that the function values at different points are correlated, where the covariance between two function values, $f(x_p)$ and $f(x_q)$, depends on the input locations, $x_p$ and $x_q$. This dependency can be specified via an arbitrary covariance function, or kernel $k(x_p, x_q)$. The most widely used kernel function is the squared exponential, or Gaussian, Kernel: $k(x_p, x_q) = \sigma_f^2 exp(\frac{-1}{2l^2}|x_p - x_q|^2)$, where $l$ is is the length scale that determines how strongly the correlation between points drops off.

Building a GP estimator still requires constructing a fingerprint, though a less sparse one. This fingerprint is used to estimate the model parameters ($l$ , $\sigma_n^2$, and $\sigma_f^2$) and to compute $f(x_*)$ for any location $x_*$.

This reduces the size of the fingerprint and provides a way for extending a sparse fingerprint to a more dense one as it gives the fingerprint values at any arbitrary location based on the assumed model. However, this comes at the cost of substantial increase in computational requirements, as we quantify in Section IV, and there is no actual saving of fingerprinting overhead as war driving has to be done to collect the training samples ($D$) anyway. Moreover, the assumed model may not fit the real environment, thus reducing the accuracy of the returned location.

Our proposed probabilistic fingerprinting technique provides accuracy better than all the current techniques with a minimal computational requirements as we quantify in Section IV.

## III. CellSense Approach

In this section, we describe our *CellSense* system for GSM phones localization. We start by an overview of the system followed by the details of the offline training and online tracking phases.

### A. Overview

*CellSense* works in two phases: an offline fingerprint construction phase and and online tracking phase. During the offline phase, a probabilistic fingerprint is constructed, where the RSSI histogram for each cell tower at a certain location is estimated. During the online tracking phase, the fingerprint is used to calculate the probability of receiving the RSSI signal strength vector at the unknown location at each location in the fingerprint. The most probable location is used as the estimated locations.

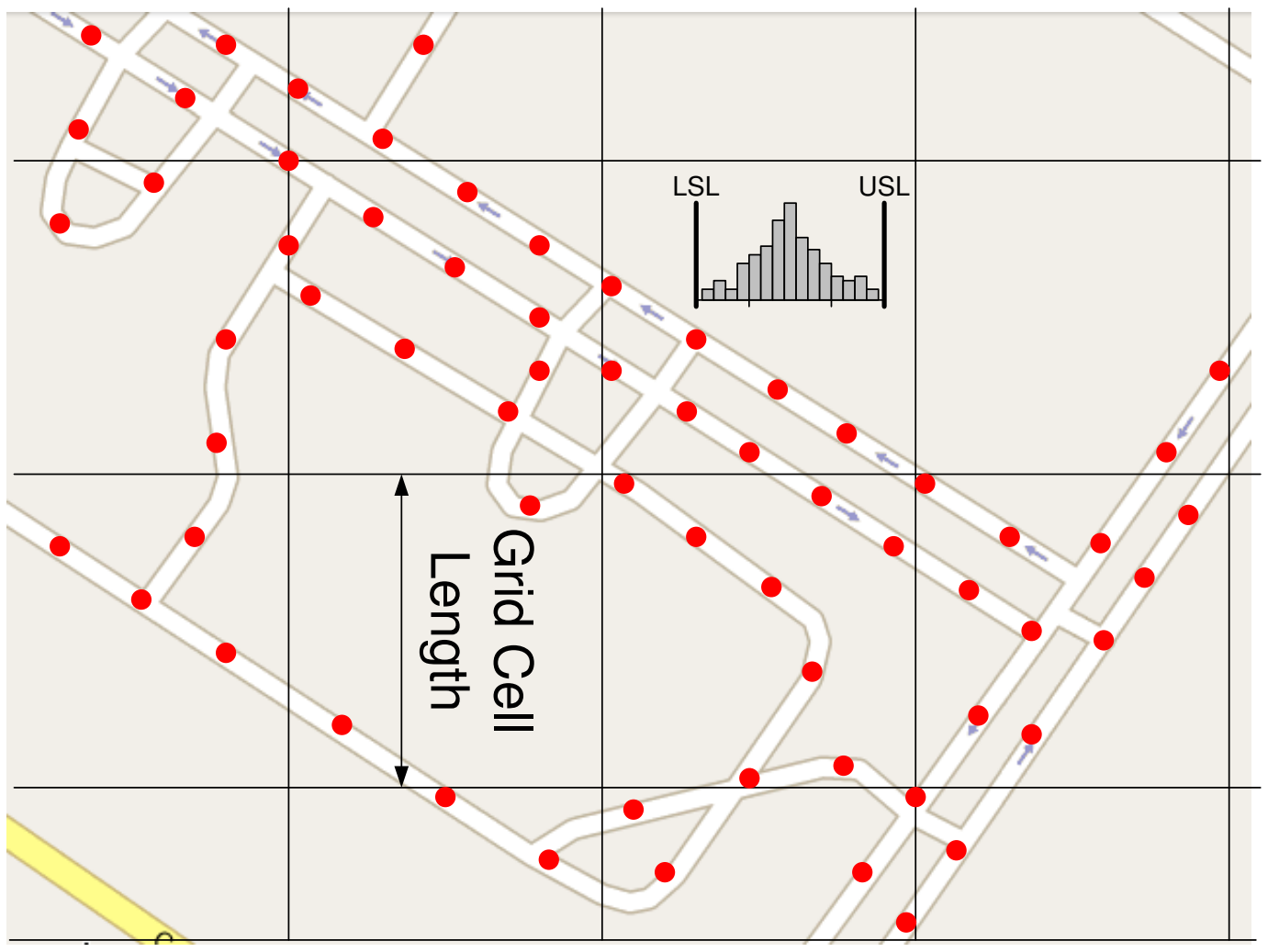

Figure 1. *CellSense* approach for fingerprint construction. The area of interest is divided into grids and the histogram is constructed using the fingerprint locations inside the grid cell. No extra overhead is required for fingerprint construction. The grid cell length parameter can be used to tradeoff accuracy and scalability.

### B. Mathematical Model

Without loss of generality, let $\mathbb{L}$ be a two dimensional physical space. Let $q$ represent the total number of cell towers in the system. We denote the $q$-dimensional signal strength space as $\mathbb{Q}$. Each element in this space is a $q$-dimensional vector whose entries represent the RSSI readings from a different cell tower. We refer to this vector as $s$. We also assume that the samples from different towers are independent. Therefore, the problem becomes, given an RSSI vector $s = (s_1, ..., s_q)$, we want to find the location $l \in \mathbb{L}$ that maximizes the probability $P(l|s)$.

### C. Offline Phase

The purpose of this phase is to construct the signal strength histogram for the RSSI received from each cell tower at each location in the fingerprint. Typically, this requires the user to stand at each location in the fingerprint for a certain time to collect enough samples to construct the RSSI histogram. This will increase the fingerprint construction overhead significantly, as the war-driving car has to stop at each location in the fingerprint for a certain time.

To avoid this overhead, we use a gridding approach, where the war-driving process is performed normally and the area of interest in divided in cells. The histogram is then constructed for each cell tower in a given cell using all fingerprint locations inside the cell, rather than for each fingerprint point (Figure 1). Note that this gridding approach reduces the resolution of the fingerprint from individual points to cells with a certain size. This not only removes the extra overhead of war-driving, but also increases the scalability of *CellSense* as the fingerprint size can be arbitrarily reduced by increasing the cell size.

### D. Online Phase

During the online phase, the user is standing at an unknown location $l$ receiving a signal strength vector $s = (s_1, ..., s_q)$, containing one entry for each cell tower. We want to find the location in the fingerprint ($l \in \mathbb{L}$) that has the maximum probability given the received signal strength vector $s$. That is, we want to find

$$argmax_l[P(l|s)] \tag{1}$$

Using Bayes' theorem, this can be written as:

$$argmax_l[P(l|s)] = argmax_l[P(s|l).\frac{P(l)}{P(s)}] \tag{2}$$

Assuming that all locations are equally probable[1] and removing $P(s)$ as it is constant for all locations, Equation 2 yields:

$$argmax_l[P(l|s)] = argmax_l[P(s|l)] \tag{3}$$

$P(s|x)$ can be calculated using the histograms constructed during the offline phase as:

$$P(s|l) = \prod_{i=1}^{q} P(s_i|l) \tag{4}$$

The above equation considers only one sample from each stream for a location estimate. In general, a number of successive samples, $N$, from each stream can be used to improve performance.

In this case, $P(s|l)$ can then be expressed as follows:

$$P(s|l) = \prod_{i=1}^{q}\prod_{j=1}^{N} P(s_{i,j}|l) \tag{5}$$

Where $s_{i,j}$ represents the $j^{th}$ sample from the $i^{th}$ stream. Thus, given the signal strength vector $s$, the discrete space estimator applies Equation 5 to calculate $P(s|l)$ for each location $l$ and returns the location that has the maximum probability.

Similarly, instead of returning just the most probable location, a weighted average of the $K$ most probable fingerprint locations, weighted by the probability of each location, can be used to obtain a better estimate of location, especially when the user is not standing exactly on a fingerprint location. We study the effect of the parameter $K$ on performance in the next section.

## IV. Performance Evaluation

In this section, we study the effect of different parameters on *CellSense* and compare its performance to other RSSI-based GSM localization systems.

[1]If the probability of being at each location is known, this can be used in the equation as is.

| Testbed | Area covered | Training set size | Testing set size | Avg. num. towers/loc. |
|---|---|---|---|---|
| One (Rural) | 1.958Km$^2$ | 1198 | 301 | 5.16 |
| Two (Urban) | 5.451Km$^2$ | 2890 | 1051 | 5.97 |

Table I
COMPARISON BETWEEN THE TWO TESTBEDS.

### A. Data Collection

We collected data for two different testbeds. The first testbed covers the Smart Village in Cairo, Egypt which represents a typical rural area. The second testbed covers a 5.5 Km$^2$ in Alexandria representing a typical urban area. Data was collected using a T-Mobile G1 phone which has a GPS receiver (used as ground truth for location) and running the Android 1.6 operating system.

We implemented the scanning program using the Android SDK. The program records the (cell-ID, signal strength, GPS location, timestamp) for the cell tower the mobile is connected to as well as the other six neighboring cell towers information as dedicated by the GSM specifications. The scanning rate was set to one per second. Two independent data sets were collected for each testbed: one for training and the other for testing. Table I summarizes the two testbeds.

### B. Effect of Changing Parameters

In this section we explore the results of changing the different parameters on the performance of *CellSense*, mainly: grid cell size, number of samples used in estimation $N$, and the number of most probable locations averaged to obtain the final location $K$.

*1) Effect of number of averaged fingerprint locations:* Figure 2 shows the effect of changing the number of the most probable locations averaged $K$ on the median localization error. The other parameters are fixed at cell size= $20m$ and $N = 1$. The figure shows that as $K$ increases, the accuracy increases. This introduces negligible increase in latency. Therefore using the center of mass of all locations as an estimate produces the best results.

*2) Effect of grid cell size:* Figure 3 shows the effect of changing the grid cell size on the median localization error. Each cell is a square with size as indicated on the x-axis. The other parameters are fixed at $N = 1$ and $K =$ all locations. The figure shows that as expected, as the cell size increases, the accuracy decreases. The figure also shows that a grid cell size up to 400 $m^2$ gives comparable accuracy to very small cell sizes for both testbeds. This indicates that *CellSense* can lead to good scalability with minimal reduction in accuracy. Moreover, the figure shows that the accuracy in urban areas is more than the accuracy in rural areas due to the increased cell tower density.

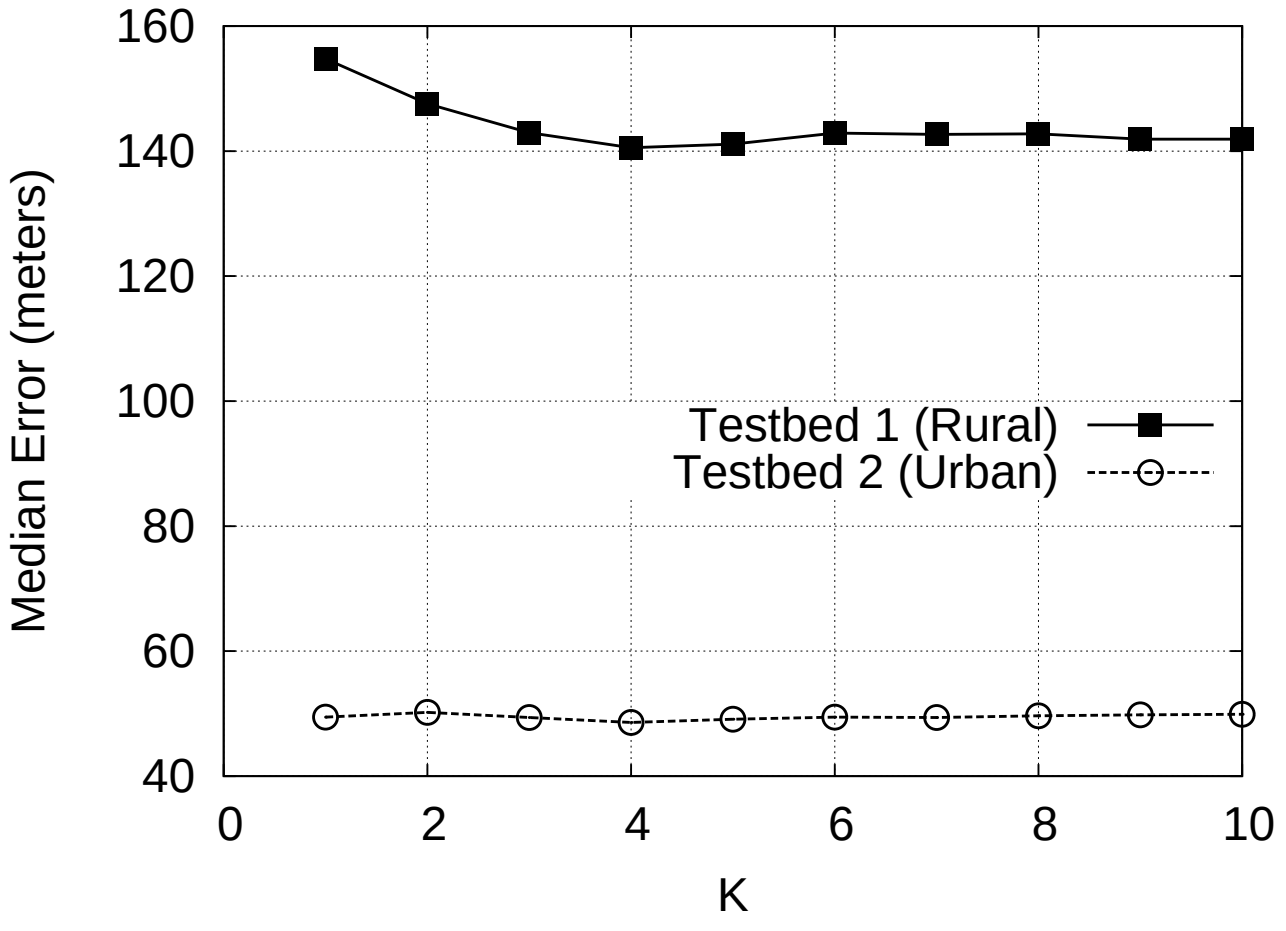

Figure 2. Effect of changing the number of most probable locations averaged ($K$) on *CellSense*'s median error.

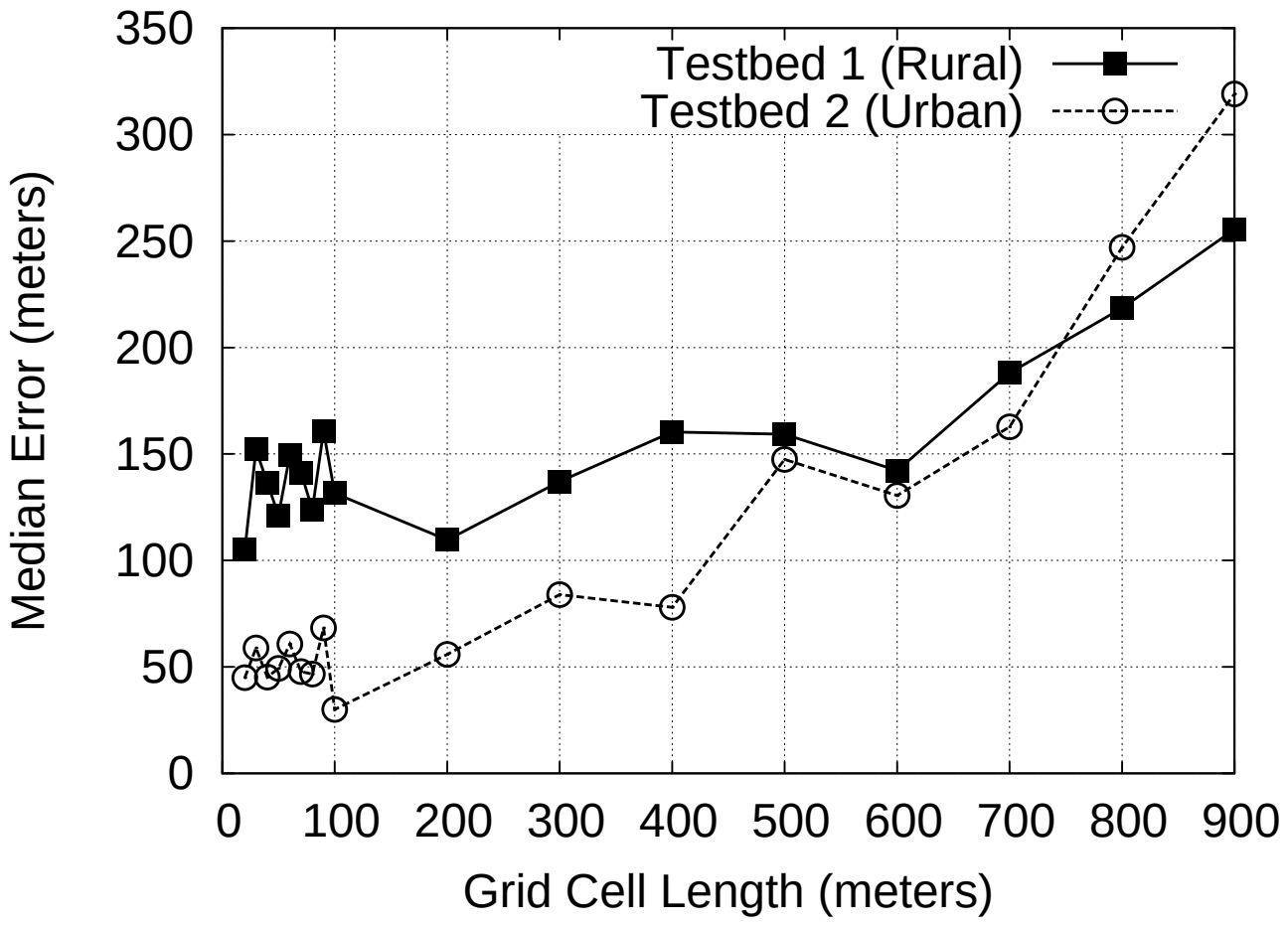

Figure 3. Effect of changing the grid cell length on *CellSense*'s median error.

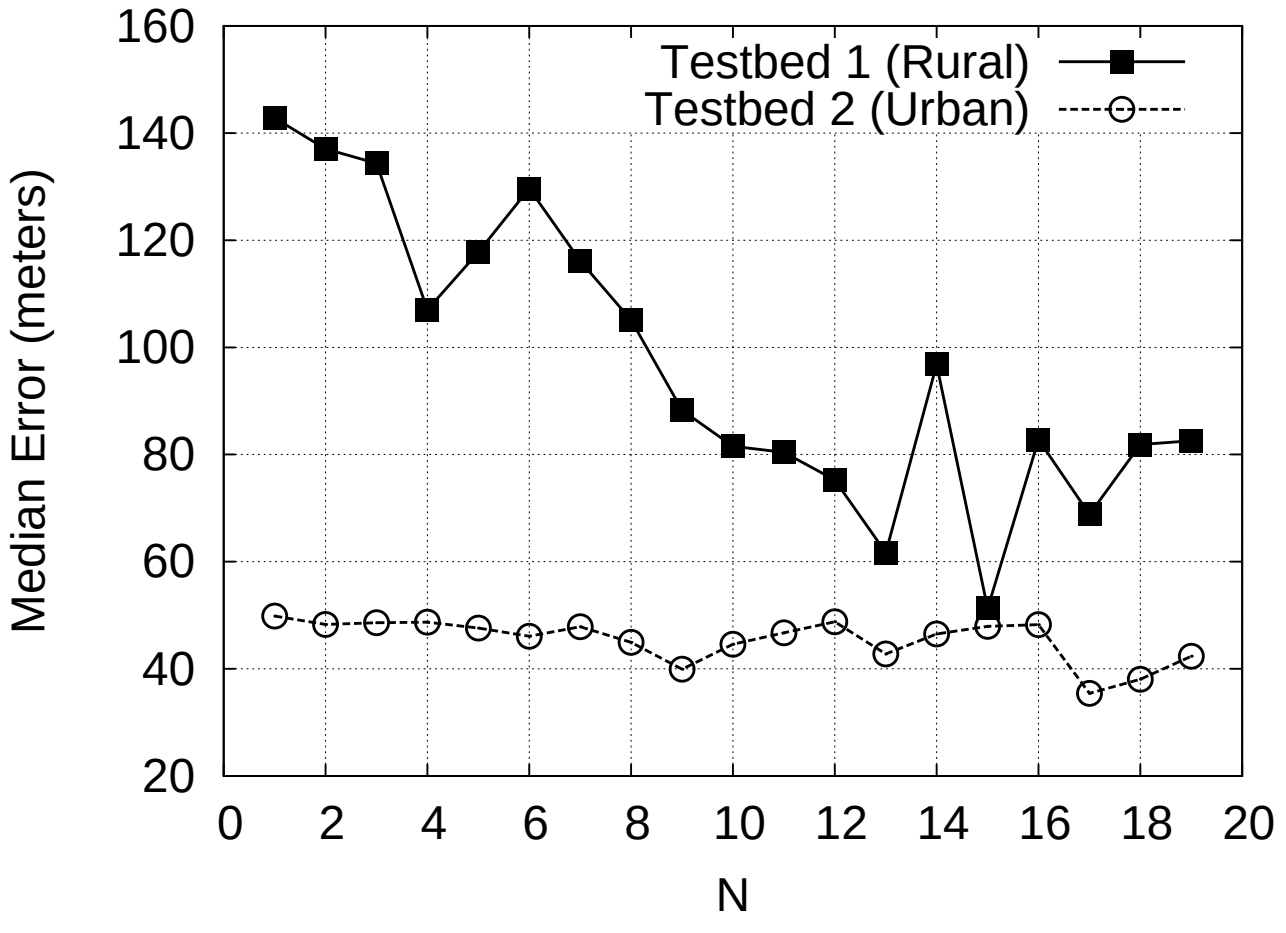

Figure 4. Effect of changing the number of samples ($N$) on *CellSense*'s median error.

*3) Effect of number of samples used $N$:* Figure 4 shows the effect of changing the number of samples used in estimation ($N$) on the median localization error. The other parameters are fixed at cell size= $20m$ and $K$ = all locations. The figure shows that as the number of samples used in estimation increases, the accuracy increases. However, the latency of obtaining a location estimate increases linearly with the number of samples used as we have to wait for these samples to be collected. Therefore, we have a tradeoff between latency and accuracy.

### *C. Comparison with Other Techniques*

In this section, we compare the performance of *CellSense*, in terms of running time and localization error, to other RSSI-based GSM localization techniques described in Section II.

*1) Localization Error:* Figure 5 shows the CDF of distance error for the different algorithms for the two testbeds. The parameters that give the best median error were used for **all** algorithms. Table II summarizes the results. The table shows that *CellSense*'s accuracy is better than any technique with at least 23.8% in rural areas and at least 86.4% in urban areas. All techniques perform better in urban areas than rural areas due to the higher density of cell towers and the more differentiation between fingerprint locations due to the dense urban area structures. Gaussian processes has better performance in urban areas than the deterministic fingerprinting technique, indicating that the Gaussian process can model the relation between signal and location relatively accurately. On the other hand, in rural environments, the performance of the Gaussian process degrades significantly.

*2) Running time:* Figure 6 compares all algorithms in terms of the average time required for one location estimate. The cell-ID based technique, i.e. Google's MyLocation, has the lowest running time. *CellSense*'s running time (averaged over all grid sizes) is slightly worse than the deterministic fingerprinting technique, while the Gaussian processes approach is two order of magnitudes worse in terms of running time. The significant accuracy advantage of *CellSense* comes at a slight degradation in performance. In addition, the grid cell size can be increased, to reduce the computational requirements of *CellSense*, with little effect on accuracy as we discussed in Section IV-B2.

## V. Conclusion

We proposed *CellSense*, a probabilistic RSSI-based fingerprinting approach for GSM cell phones. We presented the details of the system and how it constructs the probabilistic fingerprint without incurring any additional overhead. We also implemented our system on Android-based phones and compared it to other GSM-localization systems under two different testbeds. Our results show that *CellSense*'s accuracy is better than other techniques with at least 23.8% in rural areas and at least 86.4% in urban areas. This comes with a minimal increase in

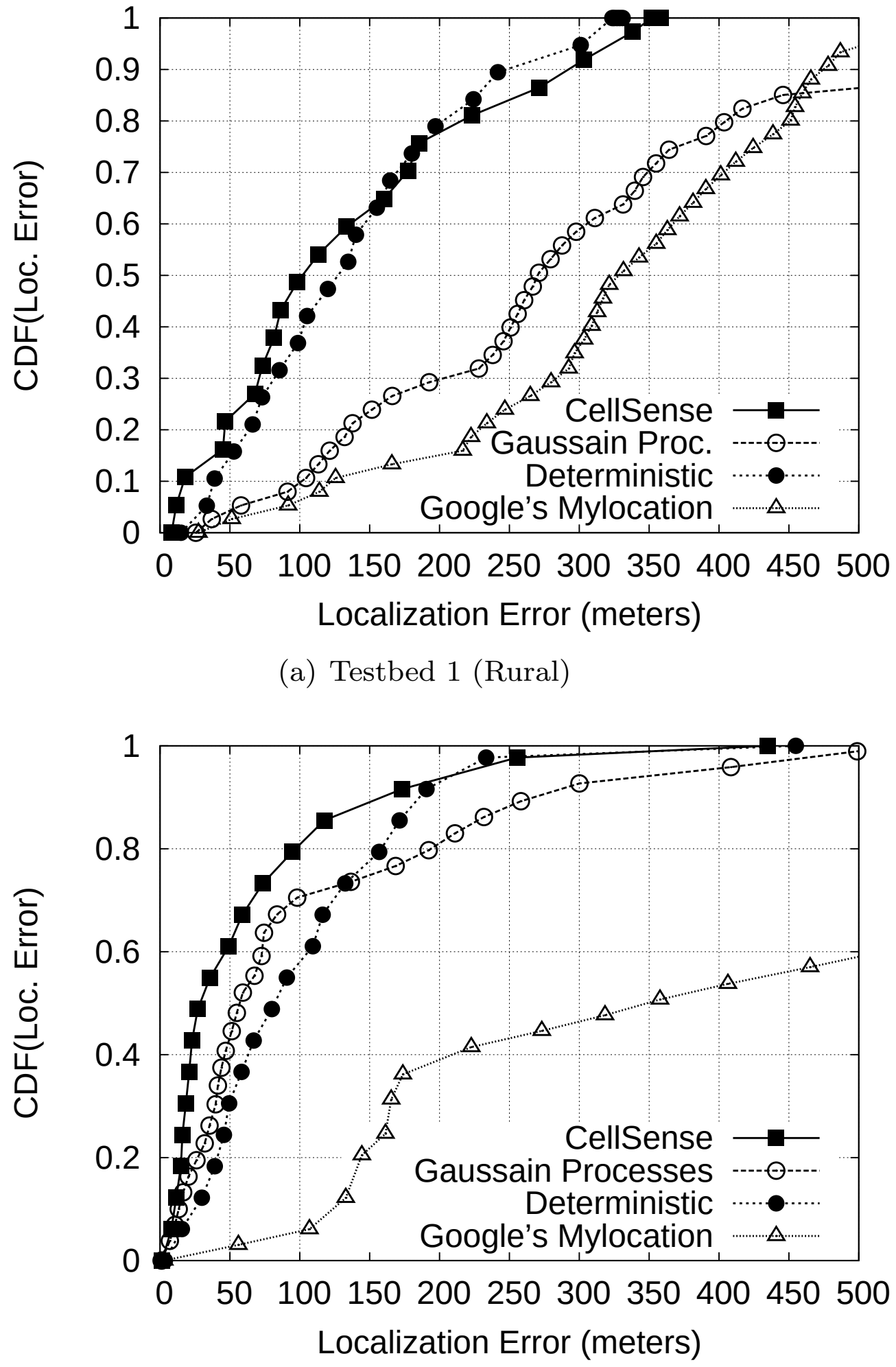

(a) Testbed 1 (Rural)

(b) Testbed 2 (Urban)

Figure 5. CDF's of distance error for different techniques under the two testbeds. CDF's for MyLocation and Gaussian Processes have been truncated.

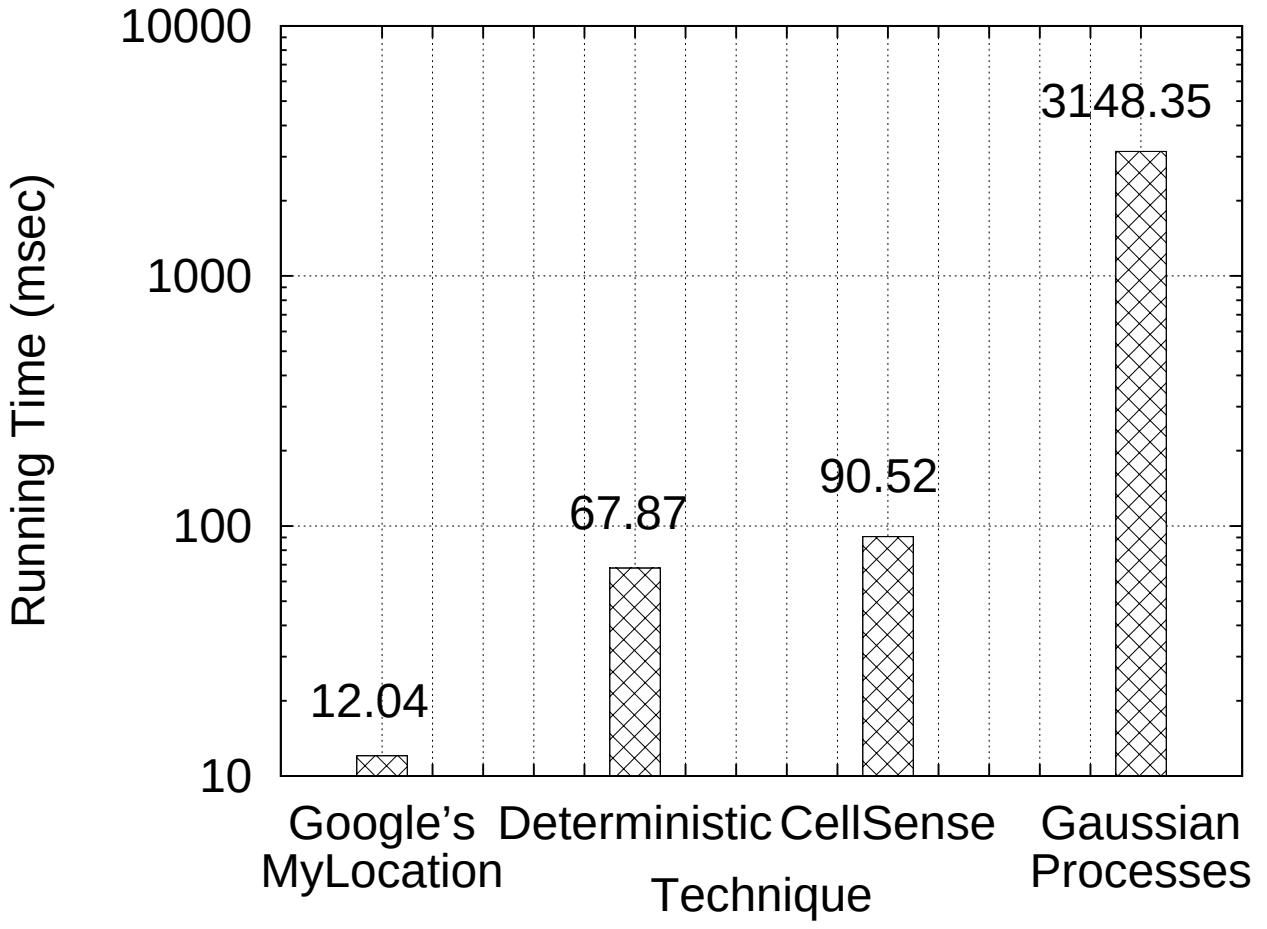

Figure 6. Running time for different techniques under the two testbeds (log scale).

| Algorithms | Google's MyLocation | Deterministic | Gaussian Processes | *CellSense* |
|---|---|---|---|---|
| Testbed 1-Rural Median Error(meters) | 327.06 (211.4%) | 130.57 (23.8%) | 270.60 (157.1%) | 105.11 |
| Testbed 2-Urban Median Error(meters) | 354.38 (1081.25%) | 89.41 (197.5%) | 56.01 (86.4%) | 30.05 |
| Average Running Time(msec) | 12.04 | 67.87 | 3148.35 | 90.52 |

Table II
COMPARISON BETWEEN DIFFERENT TECHNIQUES USING THE TWO TESTBEDS. NUMBERS BETWEEN PARENTHESIS REPRESENT PERCENTAGE DEGRADATION COMPARED TO *CellSense*.

computational requirements compared to deterministic techniques. We also studied the effect of different parameters on the accuracy-complexity tradeoff.

Currently, we are working on extending our system in different directions including using parametric distributions, clustering of fingerprint locations, experimenting with larger datasets, among others.

## ACKNOWLEDGMENT

This work is supported in part by a Google Research Award.